# Tourism Demand Forecasting: An Ensemble Deep Learning Approach

Shaolong Sun[a], Yanzhao Li[a], Ju-e Guo[a], Shouyang Wang[b,c,d,*]

[a]School of Management, Xi'an Jiaotong University, Xi'an 710049, China
[b]Academy of Mathematics and Systems Science, Chinese Academy of Sciences, Beijing 100190, China
[c]School of Economics and Management, University of Chinese Academy of Sciences, Beijing 100190, China
[d]Center for Forecasting Science, Chinese Academy of Sciences, Beijing 100190, China

*Corresponding author. Academy of Mathematics and Systems Science, Chinese Academy of Sciences, Beijing 100190, China. Tel.: +86 10 82541772; Fax: +86 10 82541972.
E-mail address: sywang@amss.ac.cn (S. Y. Wang).

**Abstract**

The availability of tourism-related big data increases the potential to improve the accuracy of tourism demand forecasting, but presents significant challenges for forecasting, including curse of dimensionality and high model complexity. A novel bagging-based multivariate ensemble deep learning approach integrating stacked autoencoders and kernel-based extreme learning machines (B-SAKE) is proposed to address these challenges in this study. By using historical tourist arrival data, economic variable data and search intensity index (SII) data, we forecast tourist arrivals in Beijing from four countries. The consistent results of multiple schemes suggest that our proposed B-SAKE approach outperforms benchmark models in terms of level accuracy, directional accuracy and even statistical significance. Both bagging and stacked autoencoder can effectively alleviate the challenges brought by tourism big data and improve the forecasting performance of the models. The ensemble deep learning model we propose contributes to tourism forecasting literature and benefits relevant government officials and tourism practitioners.

# 1. Introduction

Tourism demand forecasting plays a crucial role in tourism management. On the one hand, tourism resource planning based on accurate demand forecasting are of great significance for avoiding unnecessary losses, due to the perishability of tourism products (Chu, 2011; Law, Li, Fong, & Han, 2019; Shen, Li, & Song, 2008). On the other hand, tourism demand forecasting can effectively help the government and tourism practitioners guide tourists, and thus improve the service quality and tourists' experience (C. Zhang, Wang, Sun, & Wei, 2020).

Tourist demand forecasting approaches, used by the majority of quantitative studies, include various time series, econometric, and artificial intelligence (AI) approaches, and the combinations of these approaches (Song & Li, 2008; Song, Qiu, & Park, 2019). AI approaches can not only effectively capture the nonlinear characteristics between variables, but also require little specialized expertise as data-driven approaches. Therefore, more and more researchers develop various powerful AI approaches to further advance the literature of tourism demand forecasting (Song et al., 2019; C. Zhang et al., 2020). Notably, deep learning has been the research hotspot in this field recently (Law et al., 2019; Lv, Peng, & Wang, 2018; Y. Zhang, Li, Muskat, & Law, 2020).

The availability of tourism big data is improving gradually. In addition to historical tourist arrival data, the data widely used in current tourism demand forecasting literature mainly include economic variable data and search intensity index (SII) data. The causal econometric approaches have revealed the most crucial economic variables that determine the demand for international tourism. Specifically, these economic variables are summarized by Athanasopoulos, Song, and Sun (2017) including tourism prices in a destination relative to those in the origin country, tourism prices in competing destinations, tourists' income, and exchange rates. Additionally, with booming in Web search technology, tourists seek travel information by using search engines before traveling. These search behaviors are statistically generated into search intensity index data that could be used to accurately measure tourists' attention (Bangwayo-Skeete & Skeete, 2015; Fesenmaier, Xiang, Pan, & Law, 2010). A set of effective methods for keyword selection and data aggregation to form the indicator has been gradually established (Xin Li, Pan, Law, & Huang, 2017; Yang, Pan, Evans, & Lv, 2015). Obviously, the application techniques of SII data have been preliminarily established.

According to the concept of "data-intensive forecasting" proposed by Bunn (1989), a way to further improve forecast accuracy is by make use of the availability of multiple information and computing resources. Analogously, Song, Gao, and Lin (2013) point out that combining forecasts considering different data has become one of the most important and effective ways to improve forecasting performance. Inspired by this modeling idea, this study incorporates historical tourist arrival data, economic variable data and SII data mentioned above into the forecasting framework. Nevertheless, introducing large amounts of data also poses huge challenges for forecasting. First, tourism-related big data means many influential factors potentially affecting tourism demand. With the increase of potential features, sample data will become sparse in the feature space, which eventually leads to curse of dimensionality and affects the forecasting effect (Law et al., 2019). Second, the data with many explanatory variables also increases the complexity of the model, resulting in large variance and overfitting (Y. Zhang et al., 2020). Feature engineering is an effective way to solve the above two problems, but traditional feature extraction requires a lot of expert knowledge and manual work (Lv et al., 2018).

To address these challenges, a bagging-based multivariate ensemble deep learning model, integrating stacked autoencoders and kernel-based extreme learning machines (B-SAKE) is proposed for tourism demand forecasting. Concretely, deep learning technique automatically extracts mining features by simulating the brain's pattern to process information and does not require much domain knowledge and human resource (Pouyanfar et al., 2018). Researchers have examined unsupervised feature learning applied in tourism demand forecasting. For example, S. Li, Chen, Wang, and Ming (2018) utilize principal component analysis (PCA) to reduce the dimension of data features, thus effectively reducing the redundant information of data. Stacked autoencoder is capable of learning nonlinear relationships, which could be regarded as a more powerful nonlinear generalization of PCA. Bagging generates multiple data sets for training a set of models to improve the stability of forecasting and reduce variance effectively (Athanasopoulos et al., 2017; Inoue & Kilian, 2008) and its powerful performance has been demonstrated in many forecasting fields. Kernel-based extreme learning machine (KELM) has not only high computational efficiency, but also better forecasting performance than extreme learning machine (ELM) because the random map in ELM is replaced by the kernel in KELM (Sun, Wei, Tsui, & Wang, 2019).

We conduct numerical experiments for Beijing international tourist arrivals. To verify the effectiveness of the models, we consider the cases of tourist arrivals in Beijing city from four origin countries including United States, United Kingdom, Germany, and France. In addition to its 67 percent market share in United States, Google has more than 90% of the search market in the other three countries. SII data are more representative in these countries, and other data required in the model are publicly available. The results of empirical study are four-fold: (1) Both bagging and stacked autoencoder can improve the forecasting performance of the models. (2) Our proposed B-SAKE model is the most accurate in different forecasting schemes (i.e., onestep-ahead versus multistep-ahead, and in-sample versus out-of-sample) regarding the performance evaluation criteria including MAPE, NRMSE, and DS. (3) This forecasting model performs better than other benchmark models from the statistical perspective (DM test and PT test). (4) The consistency of our findings across four countries we considered is encouraging.

The objective of this study is to propose a novel ensemble deep learning model to mitigate the curse of dimensionality and high model complexity caused by tourism big data and verify its good forecasting accuracy and stability. The most relevant literature is Y. Zhang et al. (2020), who also point out that there may be overfitting problems in the forecasting model. They increase the data volume available for training through decomposition method and improve the efficiency of feature extraction by designing duo attention layer. Correspondingly, In the ensemble model we develop, Bagging and Stacked autoencoder are responsible for implementing similar functions. In view of excellent forecasting performance and consistency in multiple forecasting cases, our proposed B-SAKE model contributes to tourism forecasting literature.

The rest of this study is organized as follows. **Section 2** details the literature on tourism demand forecasting with search intensity index data and tourism demand forecasting with deep learning. **Section 3** introduces related methods and describes the conceptual framework of this study. **Section 4** provides a case study on Beijing tourist arrivals and compares the results of our proposed B-SAKE model with those of benchmark models. Finally, conclusions and managerial implications are summarized in **Section 5**.

## 2. Literature review

### 2.1 Tourism demand forecasting with search intensity index data

Search engines can provide a time series index of the volume of queries users using the search engine in a specific geographic area (Choi & Varian, 2012; Padhi & Pati, 2017), which is referred to in the literature as search intensity index (SII) data. Social psychologists outline the spatiotemporal frequency of specific search terms provided by web search engines can reflect the attention of specific user groups on this issue in a specific time-space (Lai, Lee, Chen, & Yu, 2017). In terms of tourism, travelers seek relevant information through search engines regarding almost all aspects of the trip, including accommodations, transportation, attractions, and dining (Fesenmaier et al., 2010; Yang et al., 2015). Therefore, SII data, as a measure of tourists' attention, has been widely used in tourism demand forecasting literature.

Choi and Varian (2012) first introduce Google Trends data to forecast visitor arrivals in Hong Kong, and the positive effect of Google Trends data in forecasting is demonstrated by using visitor arrival data from nine origin countries. However, they only consider the Google Trends index for "Vacation Destinations/Hong Kong", and the way they aggregate the data results in information loss. Given these problems, Bangwayo-Skeete and Skeete (2015) propose a novel indicator for tourism demand forecasting for countries in the Caribbean, which is based on a composite search for "hotels and flights". Yang et al. (2015) suggest that localized SII data should be selected by comparing the fitness and forecasting ability of Google Trends with those of Baidu Index. The systemic search query and selection mechanism they develop is widely accepted by later literature. Xiaoxuan Li, Wu, Peng, and Lv (2016) consider the noise contained in SII data and tourism volumes and propose a forecasting model with denoising, namely CLSI-HHT (Hibert-Huang Transform), on the basis of composite leading search index (CLSI) given by Liu, Chen, Wu, Peng, and Lv (2015) and Hibert-Huang Transform (HHT). The results demonstrate that the forecasting performance of the model without denoising is close to that of the time series model, while the CLSI-HHT model outperforms the baselines significantly. Xin Li et al. (2017) focus on SII data aggregation methods in the context of a large number of studies incorporating increasingly web search keywords. They adopt a generalized dynamic factor model (GDFM) to process many keyword variables and the proposed method improves the forecast accuracy over those of two benchmark models: a traditional time series model and a model with an index created by PCA. In recent years, some scholars have paid attention to spurious patterns in Google Trends data, such as changes in search behavior and total search volume (Bokelmann & Lessmann, 2019), and the language and platform biases that inevitably result from using SII data (Dergiades, Mavragani, & Pan, 2018), and they present corresponding improvement measures.

**2.2 Tourism demand forecasting with deep learning**

AI models have achieved successful applications in tourism demand forecasting. However, the vast majority of the AI models covered in the forecasting literature are shallow architectures, which have limited capabilities for exploring higher nonlinearities, particularly when the data have large-scale and unclear patterns (Lv et al., 2018; Zhao, Li, & Yu, 2017). In recent years, the few literatures that develop different deep learning methods have shown their great potential in improving tourism demand forecasting performance.

Lv et al. (2018) propose a novel deep learning method called the stacked autoencoder with echo-state regression (SAEN) to forecast the tourism demand. The proposed SAEN is applied in four different but representative tourism cases and the forecasting results show that SAEN outperforms the benchmark models, including SARIMA, MLR, SLFN, SVR, ESN and LSTM. Law et al. (2019) come up with two challenges (feature engineering and lag order selection) that tourism demand forecasting may face when large amounts of search engine data are adopted. And a deep network architecture for tourism demand forecasting based on LSTM is given, which not only overcomes the two challenges mentioned above, but also significantly excels support vector regression (SVR) and artificial neural network (ANN) models. Y. Zhang et al. (2020) alleviate the overfitting issue and improve tourism demand forecasting by introducing the decomposition and improving the attention mechanism. In conclusions, the above deep learning tourism demand forecasting literatures are generally aware of the curse of dimensionality and high model complexity brought by tourism big data, and have designed various forecasting frameworks to reduce possible overfitting and improve the forecasting accuracy. In the context that the field is still in the stage of continuous exploration, the ensemble deep learning framework proposed in this study is a promising approach. It is worth noting that the previous tourism deep learning forecasting literatures ignore traditional explanatory variables such as economic variables. This study tries to make up for this omission and enrich the available dataset to improve the forecasting performance.

## 3. Related methods

### 3.1 Stacked Autoencoder

Stacked autoencoder (SAE), proposed by Bengioti (2006), is a successful application of the layer-wise strategy in autoencoder. Structurally, stacked autoencoder is considered as a neural network made up several layers of autoencoders. Autoencoder, where the output is expected to reconstruct the input, is a single hidden layer feedforward neural network. Autoencoder's structure is depicted in **Fig. 1**, in which $X_I$, $H$ and $X_O$ are the input, output and hidden layer vectors, respectively. In autoencoder, "encoding" refers to the transformation from $X_I$ to $H$, and "decoding" refers to the transformation from $H$ to $X_O$. Autoencoder tries to approximate an identity function to ensure the input vector $X_I$ close to the output vector $X_O$ and realize the compressed and high-level representation of $X_I$, which is the hidden layer vector $H$. Consequently, an autoencoder is chosen to extract nonlinear features.

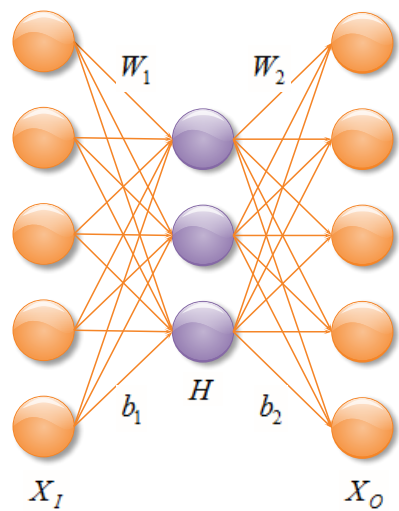

Fig. 1 The structure of an autoencoder

**Fig. 2** shows the structure of a stacked autoencoder, in which the hidden layer output of the previous autoencoder is considered to be the input of the next autoencoder. Given input data, SAE can learn effectively representations from the original input and automatically filter unrelated features. Training SAE directly on the entire structure is very time-consuming and may cause the problem of gradient disappearance, especially in the case of large network depth. Layer-wise training has always been the core of deep neural network (DNN) learning. For SAE, each autoencoder in SAE is first trained in a sequential and unsupervised manner through a back propagation (BP) algorithm. Since then, the updated parameters of SAE are shared. The SAE parameters can obtain local optimum values after layer-wise training. Moreover, SAE is not sensitive to the raw input features such that it does not require artificial feature extraction.

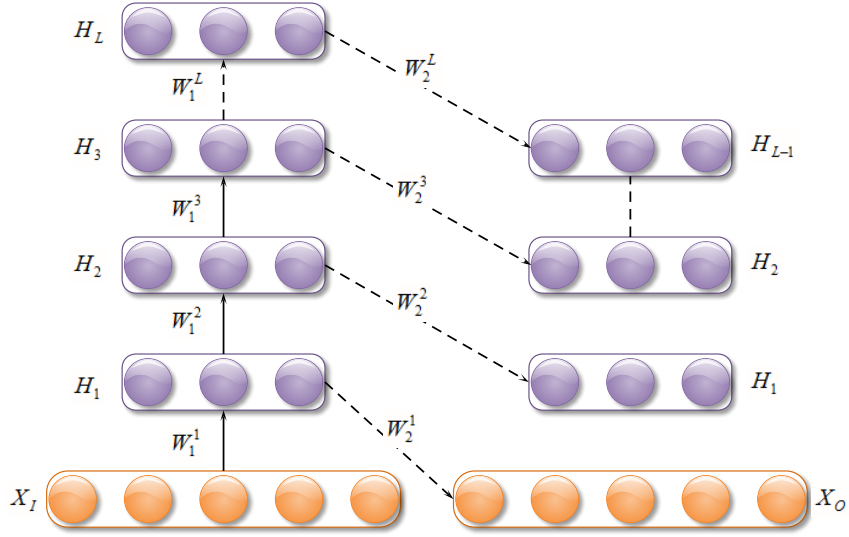

**Fig. 2** Stacked autoencoder structure

### 3.2 Kernel extreme learning machine

Extreme learning machine (ELM), proposed by Huang, Zhu, and Siew (2006), is a single-hidden layer feedforward neural network (SLFN). The ELM model has been widely used in many fields because of its high computational efficiency and generalization ability. The input weights and biases of the ELM model are randomly generated, and there is no need to adjust the hidden layer parameters. The output weights are obtained through a simple matrix computation, and this is why the ELM has high computing speed.

For $N$ samples $(x_i, y_i)$, $x_i \in \Re^N$, $y_i \in \Re^N$, $i = 1, 2, \cdots, N$. Let $h(x)$ and $Y$ be the activation function of hidden layer and the output matrix respectively. The typical ELM can be expressed as

$$Y = \begin{bmatrix} y_{1j} \\ y_{2j} \\ \vdots \\ y_{mj} \end{bmatrix}_{m \times N} = \begin{bmatrix} \sum_{i=1}^{l} \beta_{i1} h(\omega_i x_j + b_i) \\ \sum_{i=1}^{l} \beta_{i2} h(\omega_i x_j + b_i) \\ \vdots \\ \sum_{i=1}^{l} \beta_{im} h(\omega_i x_j + b_i) \end{bmatrix}_{m \times N} \quad j = 1, 2, \cdots, N \quad (1)$$

where $\omega$ is the weight between the input layer and the hidden layer, $l$ is the number of hidden layer nodes, $\beta$ represents the weight between the hidden layer and the output layer, and $b$ is the threshold of the hidden layer. The above equations can also be given by

$$H\beta = Y, \quad Y \in \Re^{N \times m}, \beta \in \Re^{N \times m}, \quad H = H(\omega, b) = h(\omega x + b) \quad (2)$$

where $H$ is the output matrix of the hidden layer. The only unknown parameter is the output weight $\beta$, which could be solved by the ordinary least squares (OLS) method. The solution of the above equation is defined as

$$\hat{\beta} = H^{\dagger}Y, \quad H^{\dagger} = H^T\left(HH^T\right)^{-1} \quad (3)$$

where $H^{\dagger}$ represents the Moore-Penrose generalized inverse of the matrix $H$. $\beta$ can be calculated by adding a positive penalty factor $1/C$ according to the theory of ridge regression and orthogonal projection method

$$\hat{\beta} = H^T\left(1/C + HH^T\right)^{-1}Y \quad (4)$$

Then the output function of the ELM could be written as

$$f(x) = H\hat{\beta} = HH^T\left(1/C + HH^T\right)^{-1}Y \quad (5)$$

This method overcomes some disadvantages of the typical gradient-based learning algorithms, such as overfitting, local minima and long computation times. The topological structure of ELM is given in **Fig. 3**.

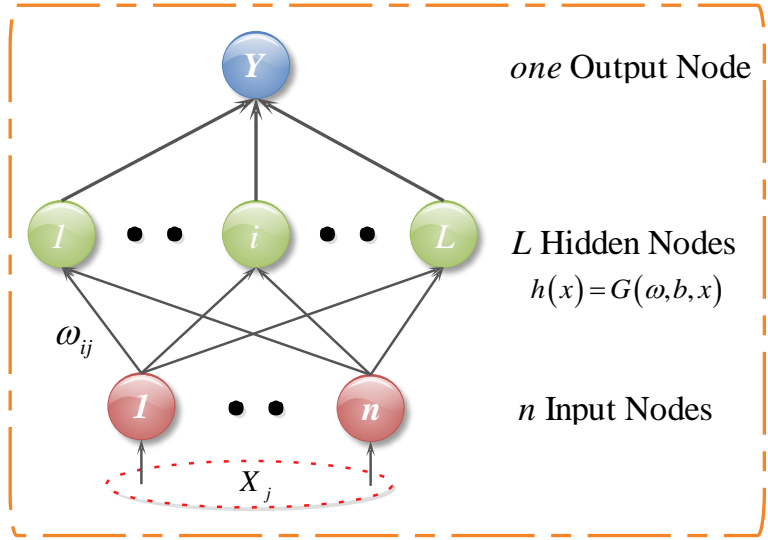

**Fig. 3** The topological structure of ELM

Huang (2014) proposed a kernel-based ELM (KELM). According to Mercer condition, the activation function $h(x)$ of the hidden layer is replaced by a kernel function. The output function of the KELM can be expresed as:

$$f(x) = h(x)\hat{\beta} = \begin{bmatrix} k(x,x_1) \\ k(x,x_2) \\ \vdots \\ k(x,x_n) \end{bmatrix}^T \left(1/C + HH^T\right)^{-1}Y \quad (6)$$

In above formula, we do not need to know the feature mapping $h(x)$, but can use its corresponding kernel function $k(x,x_i)$. This means that the kernel function can replace the random mapping of the ELM and make the output weights more stable. Thus the KELM has better generalization ability than the ELM. The flowchart of our proposed SAE-based KELM (SAKE) tourism demand forecasting approach is shown in **Fig. 4**.

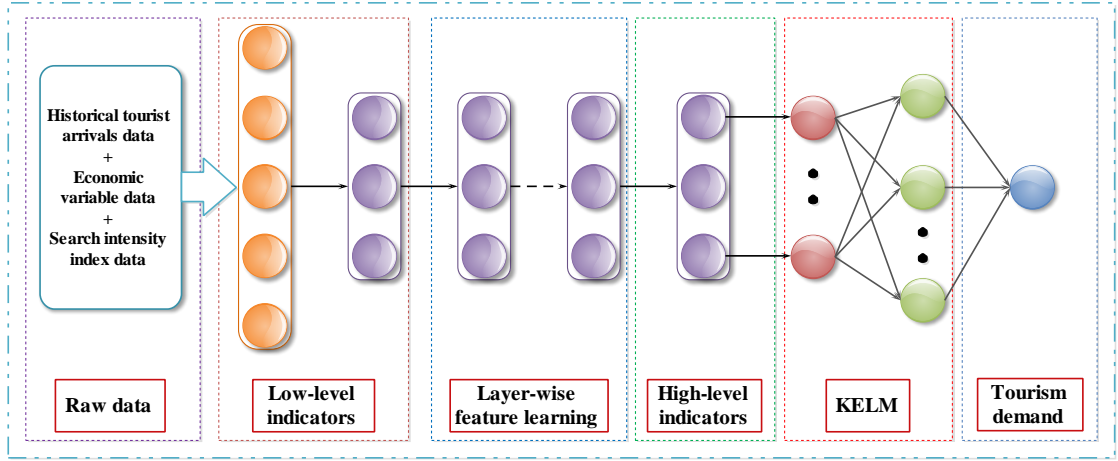

**Fig. 4** SAKE flowchart

### 3.3 Bagging

Bagging (bootstrap aggregating) is originally developed by Breiman (1996) to improve the unstable process by generating new learning sets. The purpose of bagging is to reduce the variance of forecasting and thus lead to improved accuracy. Specifically, using the resampling method, bagging generates additional samples by extracting and replacing data from the original data to train model. These additional samples are called resampling samples. We suppose that $K$ samples are generated. For each resampling sample, the process described in the previous subsection for building the SAKE network is repeated and the forecasts will be generated in each iteration. Accordingly, we now have $K$ sets of forecasts instead of one set of forecasts. In order to obtain the final forecast, we aggregate these $K$ forecasts by taking the average or the median, and the forecast variance is lower than using only one original sample.

Bagging forecasting involves generating a great number of samples, called bootstrap samples. Let $y_t$ be the predictor vector at time $t$ and $y_t = (1, l'_t)'$. Suppose $x_T$ is the latest observation where $T$ is the number of in-sample data. The in-sample data are arranged through a matrix of dimensions $(T-h) \times (1+q)$ shown as follows:

$$B = \begin{bmatrix} x_{1+h} & y'_1 \\ \vdots & \vdots \\ x_T & y'_{T-h} \end{bmatrix} \quad (7)$$

We generate a bootstrap sample $k$ by giving a replacement from the matrix $B$ blocks of $m$ rows in order to capture the dependence in the error term. It can be expressed as follows:

$$B^{(k)} = \begin{bmatrix} x^{(k)}_{1+h} & y'^{(k)}_1 \\ \vdots & \vdots \\ x^{(k)}_T & y'^{(k)}_{T-h} \end{bmatrix} \quad (8)$$

For each bootstrap sample, we perform model selection and estimate the model from matrix $B^{(k)}$ blocks. We fit the model back to the latest observations in the matrix $B$ blocks and obtain $\hat{x}_{T+h}^{(k)}$. We repeat the process for $k = 1, 2, \cdots, K$. The final forecast is then obtained as:

$$\hat{x}_{T+h}^{(bag)} = \frac{1}{K} \sum_{k=1}^{K} \hat{x}_{T+h}^{(k)} \tag{9}$$

For more details about bagging, please refer to Breiman (1996) and Athanasopoulos et al. (2017).

### 3.4 Multivariate forecasting

Multivariate forecasting, unlike univariate forecasting, takes the autoregressive effect of the target series and the impact of the exogenous variable into account. This can be denoted by

$$y(t+m) = f\left(s(y), s(x_1), \cdots, s(x_c)\right) \tag{10}$$

where $y(t+m)$ is the value of dependent variable $y$ at time $t+m$, and $s(x) = x(t), x(t-1), \cdots, x(t-N_x+1)$ past values of exogenous variable $x$ with number of $N_x$.

### 3.5 Ensemble deep learning approach

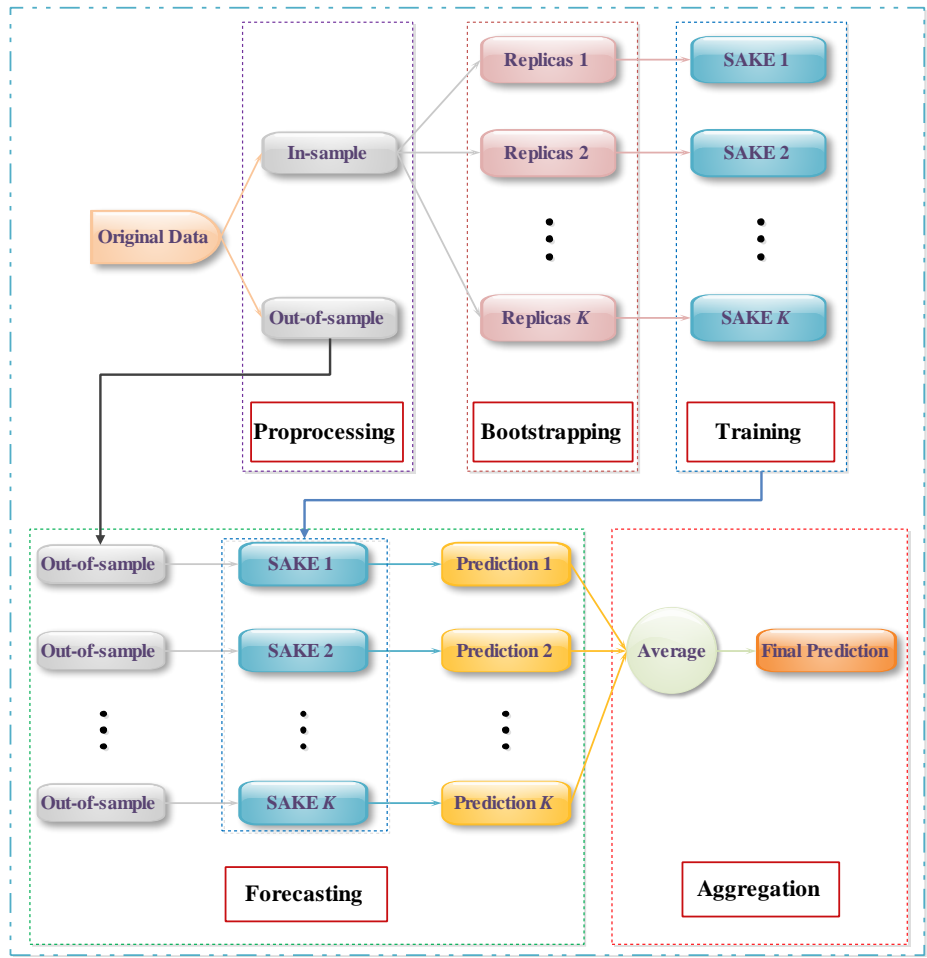

**Fig. 5** The process of B-SAKE ensemble deep learning approach

**Fig. 5** indicates the process of our proposed B-SAKE ensemble deep learning approach, which combines the advantages of bagging, SAE and KELM. This approach is composed of the following five steps:

**Data preprocessing**: transform and partition the original data into the in-sample dataset (the training set) and the out-of-sample dataset (the test set).

**Bootstrapping**: generate *K* copies of the in-sample datasets by Bagging approach.

**Model training**: train *K* SAKE models with each copy of the in-sample datasets independently.

**Individual forecasting**: generate *K* forecasts though using the *K* trained SAKE models.

**Aggregation**: take the mean value of the *K* forecasts as the final forecasting results.

## 4. Empirical study

This section provides a case study on Beijing tourist arrivals and compares the forecasting performance of our proposed B-SAKE with benchmark models. **Section 4.1** details the datasets involved in this study, including tourist arrival data, economic variable data and SII data. **Section 4.2** describes the forecasting performance evaluation criteria and statistical tests. **Section 4.3** introduces the benchmarks we choose and parameter settings. **Section 4.4** and **Section 4.5** give the forecasting results and reasonable interpretation.

### 4.1 Data

#### 4.1.1 Tourist arrival data

In this study, tourist demand (typically measured as tourist arrivals) is investigated for forecasting purposes. We select monthly inbound tourist arrivals in Beijing city over January 2008 to December 2018 from origin countries of the United States, the United Kingdom, Germany, and France, which is shown in **Fig. 6**. It is clearly observed that tourist arrivals show seasonality and volatility. The United States is the largest source of tourists for Beijing city within the four countries. The tourist arrival data are obtained from the official website of the Beijing Municipal Bureau of Statistics (http://tjj.beijing.gov.cn/), which regularly publishes monthly tourist arrivals by nationality. The datasets are divided into the in-sample dataset and the out-of-sample dataset (**see Fig. 6**). The in-sample dataset serves as model training with data from 2008.1 to 2016.12, while the out-of-sample dataset serves as model testing with data from 2017.1 to 2018.12. This division is consistent with the general laws of machine learning.

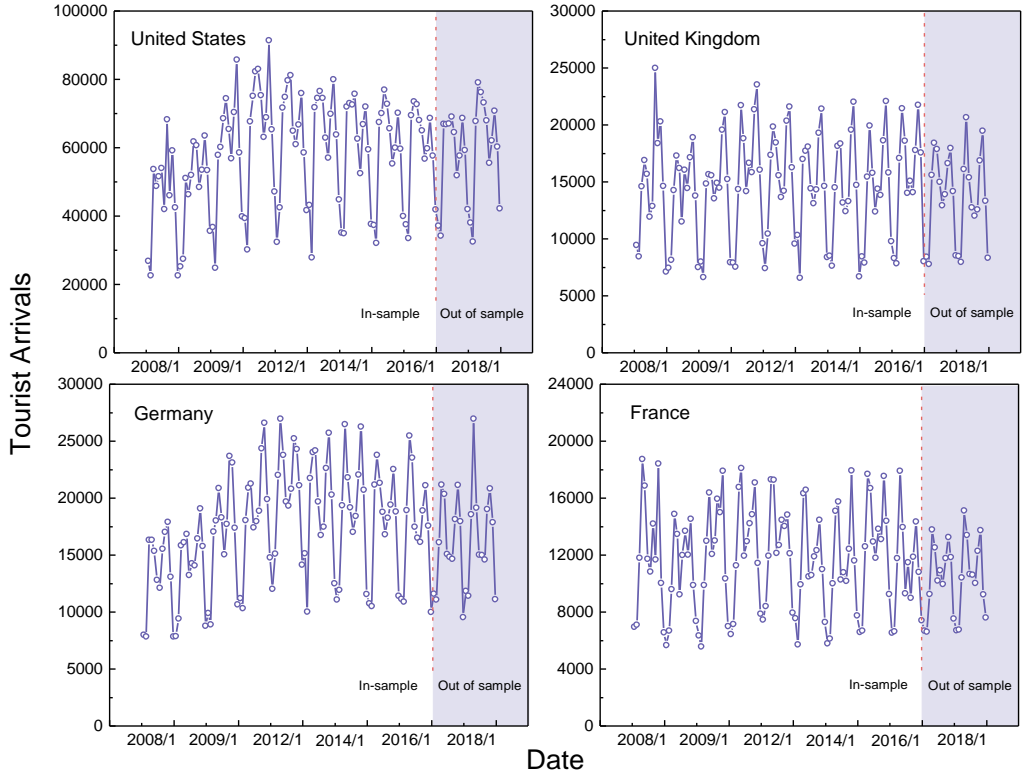

**Fig. 6** Tourist arrivals in Beijing city from four origin countries

### 4.1.2 Economic variable data

Income and price are the basic variables of economic demand theory (Crouch, 1992). Income-like and price-like variables that have been extensively validated in the international tourism demand literature (Athanasopoulos et al., 2017; G. Li, Song, & Witt, 2005; Song & Li, 2008) include (1) income level of tourists; (2) Future income expectations of tourists; (3) prices of the tourism products in the destination; (4) prices of the tourism products in the substitute destinations. It is when income and price are considered at the same time that the tourists' price perception of transnational tourism products could be accurately measured. Instead of constructing additional predictors, deep learning provides the possibility of end-to-end learning by directly utilizing typical income-like and price-like variables. This method not only greatly reduces the manual work, but also avoids the loss of forecast accuracy. Considering the availability and reliability of data, we follow Athanasopoulos et al. (2017) and choose the following economic variables.

It is expected that tourists' income level positively influences tourism demand. Tourist income level is usually measured in term of gross domestic product per capita

$$GDPpc_{i,t} \qquad (11)$$

where $i = 1, 2, \cdots, n$ represents the $n$ origin countries, $t$ is the time.

The interest rate spread reflects future economic activity and business cycle (Anderson, Athanasopoulos, & Vahid, 2007; Athanasopoulos, Hyndman, Song, & Wu, 2011; Athanasopoulos et al., 2017; Stock & Watson, 2012). Then future income expectations of tourists can be measured by the interest rate spread

$$\text{IRS}_{i,t} = LTGB_{i,t} - STGB_{i,t} \tag{12}$$

where $LTGB_{i,t}$ is the long-term government bond and $STGB_{i,t}$ is the short-term government bond of the origin country.

The demand of tourism product is inversely related to its price, which is indicated by the law of demand. The impact of exchange rates also needs to be considered for international tourism. We can use the price variable to measure this effect, which is defined as the ratio between CPIs and standardized by exchange rate

$$P_{i,t} = \frac{CPI_{CN,t} / EX_{i,t}^{CNY}}{CPI_{i,t}} \tag{13}$$

where $CPI_{i,t}$ represents the CPI of the origin country $i$ at time $t$, and $EX_{i,t}^{CNY}$ is the exchange rate between China Yuan and the currency of the origin country $i$.

Furthermore, the demand of tourism product is also affected by the prices of other tourism products. For European and American tourists, South Korea and Japan seem to be alternative options to China, and the substitute prices are defined as

$$S_{i,t}^{KR} = \frac{CPI_{KR,t}}{EX_{i,t}^{KRW}} \text{ and } S_{i,t}^{JP} = \frac{CPI_{JP,t}}{EX_{i,t}^{JPY}} \tag{14}$$

where $KRW$ and $JPY$ are Korean won and Japanese yen, respectively.

The data of economic variables mentioned above are publicly accessible and can be downloaded via *Wind* (https://www.wind.com.cn).

### 4.1.3 Search intensity index (SII) data

Following Yang et al. (2015) and Xin Li, Li, Pan, and Law (2020), we choose 24 basic search keywords in Google Trend based on the destination and various dimensions of tourism planning, including tour, lodging, recreation, traffic, dining, and shopping. The basic search keywords related to Beijing tourism are listed in **Table 1** with their corresponding dimensions. Then we search for the basic keywords in a specific origin country and set iteratively recommended keywords as the next time of search keywords. We repeat this process until there are no new keywords in the recommended list. Finally, we obtain 51 keywords, 45 keywords, 38 keywords and 33 keywords for United States, United Kingdom, Germany and France, respectively.

We calculate the Pearson correlation coefficient between tourist arrivals and keywords with different lag periods. Four correlation coefficients are calculated for each of the keyword, including the correlations between the visitor volumes in the current period and search query volumes from 1-3 months prior. We choose the keywords with the highest correlation coefficient values, which are shown in **Table 2**. To obtain the appropriate keywords, we use 0.7 as the threshold, in other words, we select the keywords with a correlation coefficient value greater than 0.7. It can be observed that the optimal lag order of most keywords is 1, indicating that tourists retrieve travel-related information one month in advance, which is consistent with our intuition.

Table 1 Basic search keywords related to Beijing tourism

| Dimension | Keywords | Dimension | Keywords | Dimension | Keywords |
|---|---|---|---|---|---|
| Tour | Beijing maps | Lodging | Beijing hotels | Recreation | Beijing bar |
| | Beijing travel | | Beijing resorts | | Beijing show |
| | Beijing weather | | Beijing restaurant | | Beijing night life |
| | Beijing travel agency | | Beijing accommodation | | Beijing recreation |
| Traffic | Beijing subway | Dining | Peking duck | Shopping | Dashilan Street |
| | Beijing flights | | Beijing food | | Panjiayuan Center |
| | Beijing airports | | Beijing snack | | Beijing shopping |
| | Beijing airlines | | Beijing food guide | | Beijing shopping guide |

Table 2 Maximum correlation coefficient of search keywords

| Countries | Keywords | Lag order | Countries | Keywords | Lag order |
|---|---|---|---|---|---|
| US | Beijing travel | 3 | UK | China travel | 2 |
| | Beijing weather | 2 | | Beijing travel | 2 |
| | China travel | 2 | | Beijing travel agency | 2 |
| | Beijing airlines | 1 | | Beijing airlines | 1 |
| | Beijing flights | 1 | | Beijing flights | 1 |
| | Beijing airports | 1 | | Beijing airports | 1 |
| | Beijing subway | 1 | | Beijing hotels | 1 |
| | Beijing hotels | 1 | | Beijing restaurant | 1 |
| | Beijing restaurant | 1 | | Peking duck | 1 |
| | Peking duck | 1 | | Duck recipes | 1 |
| | Duck recipes | 1 | | Beijing shopping | 1 |
| | Beijing shopping | 1 | | Great Wall | 1 |
| | Great Wall | 1 | | Beijing maps | 1 |
| | Forbidden city | 1 | | | |
| Germany | Beijing travel | 3 | France | Beijing tourism | 3 |
| | Beijing maps | 2 | | Beijing travel | 2 |
| | Beijing weather | 2 | | Beijing weather | 2 |
| | China travel | 2 | | Beijing flights | 1 |
| | Peking duck | 1 | | Beijing airports | 1 |
| | Beijing shopping | 1 | | Beijing hotels | 1 |
| | Great Wall | 1 | | Beijing shopping | 1 |
| | Beijing airlines | 1 | | Peking duck | 1 |
| | Beijing flights | 1 | | Great Wall | 1 |
| | Beijing airports | 1 | | Forbidden city | 1 |
| | Beijing hotels | 1 | | | |
| | Beijing restaurant | 1 | | | |

## 4.2 Performance evaluation criteria and statistic test

To evaluate and compare the forecasting performance of models, we adopt multiple error criteria commonly used in recent literature, including mean absolute percentage error (MAPE), normalized root mean square error (NRMSE) and directional symmetry (DS). The specific formulas are written as follows:

$$MAPE = \frac{1}{N}\sum_{t=1}^{N}\left|\frac{x_t - \hat{x}_t}{x_t}\right| \times 100\% \qquad (15)$$

$$NRMSE = \frac{1}{\bar{x}} \sqrt{\frac{1}{N} \sum_{t=1}^{N} (x_t - \hat{x}_t)^2} \qquad (16)$$

$$DS = \frac{1}{N-1} \sum_{t=2}^{N} d_t \times 100\%, \quad d_t = \begin{cases} 1 & if\ (x_t - x_{t-1})(\hat{x}_t - x_{t-1}) > 0 \\ 0 & otherwise \end{cases} \qquad (17)$$

where $N$ is the number of observations in the datasets, $x_t$ and $\hat{x}_t$ represent the true value and the forecasting value at time $t$, respectively. MAPE and NRMSE measure the level accuracy, the smaller the MAPE and NRMSE are, the better the level forecasting performance. DS measures the directional accuracy, the higher the DS is, the better the directional forecasting performance.

In order to exclude the influence of specific choice of data values in the sample, the DM test and PT test are employed to test the statistical significance of all models in the level forecasting and the directional forecasting, respectively. In the DM test, the null hypothesis is that the MAPE of test model is not less than that of benchmark model. The null hypothesis is rejected when DM statistics and the p-value are less than the significance level. In the PT test, the null hypothesis assumes that the true and forecast values are independently distributed. Similarly, comparing PT statistics and the corresponding p-value, the directional forecasting ability of different models can be evaluated from the statistical perspective. The process of the DM test and the PT test can be referred to Diebold and Mariano (2002) and Pesaran and Timmermann (1992).

### 4.3 Benchmarks and parameter settings

To evaluate the forecasting performance of the B-SAKE model in different forecasting schemes (onestep-ahead versus multistep-ahead, and in-sample versus out-of-sample), we formulate six benchmark models to compare with this model. ARIMA is the most widely used time-series forecasting model. Considering the seasonality and periodicity of tourism demand data, as well as the additional introduction of SII data and economic variable data, we adopt a variant of ARIMA, namely SARIMAX (S, seasonal; X, exogenous). The MLP and KELM models, as the most popular AI techniques, are widely used in tourism demand forecasting. We add stacked autoencoder network for dimension reduction on the basis of KELM to construct the SAKE model. In addition, we consider bagging-based (B-based) AI models including B-MLP, B-KELM, and B-SAKE.

The parameter specification is crucial for model performance. The appropriate parameters of SARIMAX (p,d,q)(P,D,Q,m) is estimated according to the Schwarz Criterion (SC) and Akaike Information Criterion (AIC). The numbers of hidden neurons of MLP and KELM are adjusted by trial and error experiments for minimizing in-sample forecasting errors. The Gaussian kernel function is adopted in KELM model. And the bootstrap samples of Bagging is set as 100 based on Inoue and Kilian (2008).

## 4.4 Empirical results

### 4.4.1 Forecast evaluations

We adopt the dynamic forecasting with rolling window. In the onestep-ahead forecasting, the window rolls forward one step each time, while in the multistep-ahead forecasting, the corresponding number of steps is rolled forward each time. onestep-ahead forecasting and multistep-ahead forecasting test the short-term and long-term forecasting performance of models respectively. The multistep-ahead forecasting in this study includes 3-month-ahead forecasting and 6-month-ahead forecasting. To verify the effectiveness of Bagging and SAE in dealing with overfitting and improving forecasting performance, we also designed a comparison between in-sample forecasting and out-of-sample forecasting.

We evaluate the forecasting performance of our proposed B-SAKE model and the above six benchmark models in the above forecasting schemes using the MAPE, NRMSE, and DS evaluation criteria. The results are summarized in **Tables 3-5** and it can be summarized that: (1) B-SAKE is the most accurate approach compared with the benchmark models in terms of the MAPE, NRMSE, and DS criteria. (2) The B-based models generally outperform the original models in forecast accuracy. (3) As we expected, SAKE, a SAE neural network added to KELM for dimension reduction, has better forecasting performance than KELM. (4) The SARIMAX model is the worst benchmark model, possibly because SARIMAX is not capable of efficiently capturing nonlinear patterns of tourism data in contrast to the AI models. (5) With the advance of forecasting time, the forecast accuracy of all the models decreases. (6) Although almost all model performance degrades from in-sample forecasting to out-of-sample forecasting, Bagging and SAE significantly mitigate this problem, especially in short-term forecasting.

In particular, our proposed B-SAKE model achieves the best forecast accuracy in the four origin countries, which is shown in bold in the tables. Taking the example of the United States, the reductions in MAPE are 90.89%, 87.41%, 76.70%, 75.58%, 68.92% and 34.96% in comparison with those of SARIMAX, MLP, B-MLP, KELM, B-KELM and SAKE in the case of 1-month-ahead forecasting. For NRMSE the reductions are 89.80%, 85.60%, 79.31%, 78.64%, 69.86% and 33.33%, respectively. B-SAKE achieves 77.36% better directional forecasts than SARIMAX and 20%-30% better directional forecasts than the AI models. From in-sample forecasting to out-of-sample forecasting, the accuracy loss of MAPE after applying Bagging on the SAKE decreases from 18.07% to 9.33%, compared with that of the SAKE without Bagging. Analogously, SAE helps KELM reduce the accuracy loss of MAPE from 30.61% to 18.07%. It is clearly illustrated that B-SAKE model is a highly promising forecasting approach.

Table 3 Forecasting performance of different models: 1-month-ahead forecasting

| Countries | Models | In-sample | | | Out-of-sample | | |
|---|---|---|---|---|---|---|---|
| | | MAPE | NRMSE | DS | MAPE | NRMSE | DS |

| Countries | Models | In-sample | | | Out-of-sample | | |
|---|---|---|---|---|---|---|---|
| | | MAPE | NRMSE | DS | MAPE | NRMSE | DS |
| US | SARIMAX | 5.413 | 6.019 | 55.21 | 5.935 | 6.843 | 54.17 |
| | MLP | 3.915 | 4.263 | 72.92 | 4.017 | 4.586 | 62.50 |
| | B-MLP | 2.116 | 2.968 | 78.13 | 2.873 | 3.514 | 66.67 |
| | KELM | 2.019 | 2.875 | 77.08 | 2.637 | 3.142 | 70.83 |
| | B-KELM | 1.586 | 2.037 | 79.17 | 1.781 | 2.364 | 75.00 |
| | SAKE | 0.758 | 0.921 | 86.46 | 0.895 | 1.002 | 83.33 |
| | B-SAKE | **0.493** | **0.614** | **97.92** | **0.539** | **0.684** | **100.00** |
| UK | SARIMAX | 5.321 | 5.873 | 57.29 | 5.765 | 6.521 | 58.33 |
| | MLP | 3.874 | 4.301 | 73.96 | 4.115 | 4.453 | 66.67 |
| | B-MLP | 2.106 | 2.829 | 77.08 | 2.704 | 3.437 | 70.83 |
| | KELM | 2.113 | 2.807 | 76.04 | 2.638 | 3.012 | 75.00 |
| | B-KELM | 1.602 | 2.115 | 78.13 | 1.876 | 2.364 | 79.17 |
| | SAKE | 0.927 | 1.206 | 87.50 | 1.049 | 1.258 | 87.50 |
| | B-SAKE | **0.583** | **0.705** | **95.83** | **0.639** | **0.794** | **95.83** |
| Germany | SARIMAX | 5.217 | 5.638 | 56.25 | 5.601 | 6.332 | 54.17 |
| | MLP | 3.746 | 4.105 | 75.00 | 4.015 | 4.307 | 66.67 |
| | B-MLP | 2.019 | 2.736 | 79.17 | 2.507 | 3.275 | 75.00 |
| | KELM | 2.005 | 2.693 | 78.13 | 2.439 | 2.982 | 75.00 |
| | B-KELM | 1.403 | 2.012 | 80.21 | 1.631 | 2.204 | 79.17 |
| | SAKE | 0.809 | 0.994 | 88.54 | 0.947 | 1.143 | 87.50 |
| | B-SAKE | **0.514** | **0.621** | **100.00** | **0.548** | **0.683** | **95.83** |
| France | SARIMAX | 5.447 | 5.906 | 54.17 | 5.907 | 6.492 | 54.17 |
| | MLP | 3.896 | 4.251 | 71.88 | 4.206 | 4.517 | 62.50 |
| | B-MLP | 2.258 | 2.906 | 76.04 | 2.844 | 3.352 | 66.67 |
| | KELM | 2.165 | 2.884 | 76.04 | 2.608 | 3.108 | 70.83 |
| | B-KELM | 1.652 | 2.067 | 78.13 | 1.835 | 2.216 | 75.00 |
| | SAKE | 0.944 | 1.305 | 85.42 | 1.106 | 1.295 | 79.17 |
| | B-SAKE | **0.613** | **0.745** | **93.75** | **0.701** | **0.785** | **91.67** |

Table 4 Forecasting performance of different models: 3-month-ahead forecasting

| Countries | Models | In-sample | | | Out-of-sample | | |
|---|---|---|---|---|---|---|---|
| | | MAPE | NRMSE | DS | MAPE | NRMSE | DS |
| US | SARIMAX | 5.501 | 5.943 | 53.19 | 5.942 | 6.886 | 54.17 |
| | MLP | 3.896 | 4.128 | 57.45 | 4.109 | 4.593 | 58.33 |
| | B-MLP | 2.207 | 3.146 | 60.64 | 2.894 | 3.571 | 62.50 |
| | KELM | 2.158 | 3.053 | 65.96 | 2.729 | 3.203 | 62.50 |
| | B-KELM | 1.703 | 2.214 | 70.21 | 1.834 | 2.385 | 66.67 |
| | SAKE | 0.895 | 1.102 | 79.79 | 0.912 | 1.175 | 79.17 |
| | B-SAKE | **0.584** | **0.739** | **90.43** | **0.613** | **0.794** | **87.50** |
| UK | SARIMAX | 5.408 | 5.886 | 55.32 | 5.802 | 6.613 | 54.17 |
| | MLP | 3.940 | 4.395 | 60.64 | 4.124 | 4.485 | 62.50 |
| | B-MLP | 2.251 | 3.058 | 68.09 | 2.718 | 3.445 | 66.67 |
| | KELM | 2.236 | 2.984 | 69.15 | 2.695 | 3.101 | 66.67 |
| | B-KELM | 1.742 | 2.205 | 73.40 | 1.925 | 2.397 | 70.83 |
| | SAKE | 1.025 | 1.374 | 82.98 | 1.139 | 1.402 | 79.17 |
| | B-SAKE | **0.675** | **0.793** | **91.49** | **0.725** | **0.884** | **87.50** |
| Germany | SARIMAX | 5.321 | 5.759 | 52.13 | 5.711 | 6.425 | 45.83 |
| | MLP | 3.853 | 4.256 | 59.57 | 4.096 | 4.396 | 58.33 |
| | B-MLP | 2.127 | 2.858 | 63.83 | 2.617 | 3.364 | 62.50 |
| | KELM | 2.207 | 2.801 | 68.09 | 2.545 | 3.022 | 66.67 |
| | B-KELM | 1.526 | 2.106 | 72.34 | 1.726 | 2.307 | 70.83 |
| | SAKE | 0.915 | 1.145 | 78.72 | 1.021 | 1.253 | 75.00 |
| | B-SAKE | **0.608** | **0.733** | **89.36** | **0.657** | **0.760** | **83.33** |
| France | SARIMAX | 5.514 | 6.004 | 51.06 | 6.014 | 6.583 | 50.00 |
| | MLP | 3.926 | 4.296 | 58.51 | 4.310 | 4.594 | 54.17 |
| | B-MLP | 2.338 | 2.988 | 64.89 | 2.896 | 3.412 | 58.33 |

| | KELM | 2.253 | 2.903 | 67.02 | 2.711 | 3.213 | 62.50 |
| | B-KELM | 1.751 | 2.175 | 71.28 | 1.905 | 2.305 | 70.83 |
| | SAKE | 1.031 | 1.412 | 77.66 | 1.217 | 1.309 | 75.00 |
| | B-SAKE | **0.709** | **0.816** | **88.30** | **0.819** | **0.901** | **83.33** |

Table 5 Forecasting performance of different models: 6-month-ahead forecasting

| Countries | Models | In-sample | | | Out-of-sample | | |
| --- | --- | --- | --- | --- | --- | --- | --- |
| | | MAPE | NRMSE | DS | MAPE | NRMSE | DS |
| US | SARIMAX | 6.204 | 6.913 | 51.65 | 6.585 | 6.963 | 50.00 |
| | MLP | 5.873 | 5.943 | 56.04 | 6.036 | 6.147 | 54.17 |
| | B-MLP | 4.161 | 4.207 | 59.34 | 4.543 | 4.601 | 58.33 |
| | KELM | 4.035 | 4.352 | 63.74 | 4.167 | 4.375 | 62.50 |
| | B-KELM | 3.256 | 3.106 | 69.23 | 3.402 | 3.321 | 66.67 |
| | SAKE | 1.758 | 1.701 | 78.02 | 1.808 | 1.905 | 75.00 |
| | B-SAKE | **1.142** | **1.235** | **87.91** | **1.236** | **1.343** | **83.33** |
| UK | SARIMAX | 6.385 | 6.749 | 52.75 | 6.601 | 6.835 | 54.17 |
| | MLP | 5.741 | 5.810 | 58.24 | 5.904 | 6.024 | 58.33 |
| | B-MLP | 4.359 | 4.361 | 62.64 | 4.498 | 4.535 | 62.50 |
| | KELM | 4.336 | 4.458 | 67.03 | 4.441 | 4.601 | 62.50 |
| | B-KELM | 3.104 | 3.267 | 72.53 | 3.517 | 3.358 | 66.67 |
| | SAKE | 1.837 | 1.901 | 81.32 | 1.934 | 2.043 | 79.17 |
| | B-SAKE | **1.206** | **1.348** | **89.01** | **1.319** | **1.455** | **83.33** |
| Germany | SARIMAX | 6.103 | 6.258 | 50.55 | 6.214 | 6.359 | 45.83 |
| | MLP | 5.507 | 5.654 | 57.14 | 5.639 | 5.832 | 54.17 |
| | B-MLP | 4.124 | 4.235 | 60.44 | 4.147 | 4.353 | 58.33 |
| | KELM | 4.048 | 4.209 | 64.84 | 4.265 | 4.298 | 62.50 |
| | B-KELM | 2.943 | 3.036 | 70.33 | 3.107 | 3.104 | 66.67 |
| | SAKE | 1.714 | 1.854 | 76.92 | 1.851 | 1.985 | 70.83 |
| | B-SAKE | **1.025** | **1.146** | **86.81** | **1.138** | **1.268** | **79.17** |
| France | SARIMAX | 6.269 | 6.437 | 51.65 | 6.585 | 6.731 | 45.83 |
| | MLP | 5.543 | 5.610 | 56.04 | 5.836 | 5.836 | 50.00 |
| | B-MLP | 4.301 | 4.411 | 61.54 | 4.501 | 4.602 | 54.17 |
| | KELM | 4.258 | 4.374 | 63.74 | 4.394 | 4.589 | 58.33 |
| | B-KELM | 3.043 | 3.165 | 69.23 | 3.452 | 3.293 | 58.33 |
| | SAKE | 1.804 | 1.987 | 75.82 | 1.991 | 2.120 | 70.83 |
| | B-SAKE | **1.267** | **1.359** | **85.71** | **1.368** | **1.487** | **79.17** |

### 4.4.2 Statistic tests

To further verify the level and directional forecasting performance of the B-SAKE model from the statistical perspective, the DM test and the PT test are also employed to test the statistical significance of all the models within the out-of-sample data. **Tables 6-9** report the results of the DM test and the PT test with respect to different forecasting horizons. The numbers outside the brackets in the tables are the DM statistics or PT statistics while the numbers inside the brackets are the corresponding p-values.

According to the DM test results (**Table 6-8**), we can observe that as the forecasting horizon increases, the forecasting performance of the model decreases. Even so, when testing the B-SAKE model, all the DM tests are less than -1.7013 corresponding to p-values less than 0.0444, which means that the B-SAKE model outperforms other benchmark models in Multistep-ahead forecasting scheme under the 95% confidence level. This indicates the superiority of the B-SAKE model.

Specifically, we note that when the B-SAKE is tested against the SARIMAX model, B-SAKE model statistically confirms its superiority under the 100% confidence level. Furthermore, **Tables 6-8** show that the level forecasting performance of the models increases successively for SARIMAX, MLP, B-MLP, KELM, B-KELM, SAKE and B-SAKE in multistep-ahead forecasting scheme. B-based models and SAE-based model outperform the original models under the 95% confidence level, which demonstrates the effectiveness of Bagging and SAE.

The results of PT test are displayed in **Table 9**. There is no doubt that all models' directional accuracy in long-term forecasting declines. However, the forecasting results of the B-KELM, SAKE, and B-SAKE models reject the null hypothesis in both short-term forecasting and long-term forecasting under near the 100% confidence level, which indicate the powerful performance of these three models in the directional forecasting. The B-SAKE performs the best in all schemes, followed by the SAKE, B-KELM, KELM, B-MLP, MLP, and SARIMAX. SARIMAX is almost ineffective in long-term forecasting where its p-values are greater than 0.1. Bagging and SAE can significantly improve performance in the directional forecasting.

Table 6 The DM test results for different models in 1-month-ahead forecasting

| Countries | Models | B-SAKE | SAKE | B-KELM | KELM | B-MLP | MLP |
|---|---|---|---|---|---|---|---|
| US | SAKE | -1.9873 (0.0234) | | | | | |
| | B-KELM | -2.0358 (0.0209) | -1.9461 (0.0258) | | | | |
| | KELM | -2.3879 (0.0085) | -2.0133 (0.0220) | -1.8742 (0.0305) | | | |
| | B-MLP | -2.9367 (0.0017) | -2.5381 (0.0056) | -1.8943 (0.0340) | -1.8627 (0.0291) | | |
| | MLP | -3.8749 (0.0001) | -3.4106 (0.0003) | -2.2033 (0.0014) | -2.1782 (0.0147) | -1.8658 (0.0310) | |
| | SARIMAX | -4.5037 (0.0000) | -4.2917 (0.0000) | -4.0143 (0.0000) | -3.9741 (0.0000) | -3.2859 (0.0005) | -2.9576 (0.0016) |
| UK | SAKE | -1.9560 (0.0252) | | | | | |
| | B-KELM | -1.9983 (0.0228) | -1.9562 (0.0252) | | | | |
| | KELM | -2.2749 (0.0115) | -2.1247 (0.0168) | -1.8868 (0.0296) | | | |
| | B-MLP | -2.8907 (0.0019) | -2.4359 (0.0074) | -1.8856 (0.0297) | -1.8963 (0.0290) | | |
| | MLP | -3.6358 (0.0001) | -3.4267 (0.0003) | -2.3748 (0.0088) | -2.2041 (0.0138) | -1.8742 (0.0305) | |
| | SARIMAX | -4.4937 (0.0000) | -4.2341 (0.0000) | -4.1045 (0.0000) | -3.9648 (0.0000) | -3.3058 (0.0005) | -2.8937 (0.0019) |
| Germany | SAKE | -1.9843 (0.0236) | | | | | |
| | B-KELM | -2.0687 (0.0193) | -1.9687 (0.0245) | | | | |
| | KELM | -2.2975 (0.0108) | -2.1589 (0.0154) | -1.8963 (0.0290) | | | |
| | B-MLP | -2.9954 (0.0014) | -2.3946 (0.0083) | -1.8756 (0.0304) | -1.9354 (0.0265) | | |
| | MLP | -3.7025 (0.0001) | -3.5085 (0.0002) | -2.4019 (0.0082) | -2.2241 (0.0131) | -1.8965 (0.0289) | |
| | SARIMAX | -4.5136 (0.0000) | -4.3541 (0.0000) | -4.0157 (0.0000) | -4.0156 (0.0000) | -3.4523 (0.0003) | -2.9632 (0.0015) |
| France | SAKE | -1.8994 (0.0288) | | | | | |
| | B-KELM | -1.9536 (0.0254) | -1.8713 (0.0307) | | | | |
| | KELM | -2.1254 (0.0168) | -2.0145 (0.0220) | -1.8623 (0.0313) | | | |
| | B-MLP | -2.7412 (0.0031) | -2.3657 (0.0090) | -1.7843 (0.0372) | -1.9602 (0.0250) | | |
| | MLP | -3.5896 (0.0002) | -3.3215 (0.0004) | -2.2250 (0.0130) | -2.2247 (0.0131) | -1.8690 (0.0308) | |
| | SARIMAX | -4.3582 (0.0000) | -4.1598 (0.0000) | -4.0236 (0.0000) | -3.8543 (0.0001) | -3.2968 (0.0005) | -2.8036 (0.0025) |

Table 7 The DM test results for different models in 3-month-ahead forecasting

| Countries | Models | B-SAKE | SAKE | B-KELM | KELM | B-MLP | MLP |
|---|---|---|---|---|---|---|---|
| US | SAKE | -1.9925 (0.0232) | | | | | |

|  | B-KELM | -2.1426 (0.0161) | -1.9895 (0.0233) | | | | |
|---|---|---|---|---|---|---|---|
|  | KELM | -2.2567 (0.0120) | -2.1103 (0.0174) | -1.8803 (0.0300) | | | |
|  | B-MLP | -2.8893 (0.0019) | -2.6177 (0.0044) | -1.9011 (0.0286) | -1.8511 (0.0321) | | |
|  | MLP | -3.6254 (0.0001) | -3.4526 (0.0003) | -2.0219 (0.0216) | -1.9416 (0.0261) | -1.7913 (0.0366) | |
|  | SARIMAX | -4.4029 (0.0000) | -4.2103 (0.0000) | -4.0014 (0.0000) | -3.8916 (0.0000) | -3.2341 (0.0006) | -2.6011 (0.0046) |
|  | SAKE | -1.8365 (0.0331) | | | | | |
|  | B-KELM | -1.9141 (0.0278) | -1.8854 (0.0297) | | | | |
| UK | KELM | -2.1056 (0.0176) | -2.0112 (0.0222) | -1.8103 (0.0351) | | | |
|  | B-MLP | -2.7319 (0.0031) | -2.5133 (0.0060) | -1.8995 (0.0287) | -1.8356 (0.0332) | | |
|  | MLP | -3.4103 (0.0003) | -3.2608 (0.0006) | -2.1236 (0.0169) | -2.0143 (0.0220) | -1.8041 (0.0356) | |
|  | SARIMAX | -4.3291 (0.0000) | -4.1890 (0.0000) | -4.0126 (0.0000) | -3.7859 (0.0001) | -3.2210 (0.0006) | -2.5961 (0.0047) |
|  | SAKE | -1.8511 (0.0321) | | | | | |
|  | B-KELM | -1.9954 (0.0230) | -1.8453 (0.0325) | | | | |
| Germany | KELM | -2.1063 (0.0176) | -2.0043 (0.0225) | -1.8510 (0.0321) | | | |
|  | B-MLP | -2.6917 (0.0036) | -2.3120 (0.0104) | -1.9013 (0.0286) | -1.8363 (0.0332) | | |
|  | MLP | -3.4011 (0.0003) | -3.3017 (0.0005) | -2.2163 (0.0133) | -2.1247 (0.0168) | -1.8142 (0.0348) | |
|  | SARIMAX | -4.2985 (0.0000) | -4.1066 (0.0000) | -3.9859 (0.0000) | -3.7956 (0.0001) | -3.3018 (0.0005) | -2.5385 (0.0056) |
|  | SAKE | -1.7995 (0.0360) | | | | | |
|  | B-KELM | -1.8563 (0.0317) | -1.7956 (0.0363) | | | | |
| France | KELM | -1.9863 (0.0235) | -1.9983 (0.0228) | -1.8152 (0.0347) | | | |
|  | B-MLP | -2.2106 (0.0135) | -2.1183 (0.0171) | -1.8025 (0.0357) | -1.8564 (0.0317) | | |
|  | MLP | -3.3142 (0.0005) | -3.1745 (0.0008) | -2.1195 (0.0170) | -2.0016 (0.0227) | -1.7974 (0.0361) | |
|  | SARIMAX | -4.1107 (0.0000) | -4.0135 (0.0000) | -3.8983 (0.0000) | -3.6968 (0.0001) | -3.1163 (0.0009) | -2.4101 (0.0080) |

Table 8 The DM test results for different models in 6-month-ahead forecasting

| Countries | Models | B-SAKE | SAKE | B-KELM | KELM | B-MLP | MLP |
|---|---|---|---|---|---|---|---|
|  | SAKE | -1.8013 (0.0358) | | | | | |
|  | B-KELM | -2.0527 (0.0201) | -1.8416 (0.0328) | | | | |
| US | KELM | -2.2015 (0.0139) | -2.0234 (0.0215) | -1.7983 (0.0361) | | | |
|  | B-MLP | -2.7749 (0.0028) | -2.4517 (0.0071) | -1.8546 (0.0318) | -1.7634 (0.0389) | | |
|  | MLP | -3.4135 (0.0003) | -3.2635 (0.0006) | -2.0034 (0.0226) | -1.8125 (0.0350) | -1.7011 (0.0445) | |
|  | SARIMAX | -4.2011 (0.0000) | -4.1063 (0.0000) | -3.9568 (0.0000) | -3.7412 (0.0001) | -3.0253 (0.0012) | -2.2103 (0.0135) |
|  | SAKE | -1.7413 (0.0408) | | | | | |
|  | B-KELM | -1.8041 (0.0356) | -1.7998 (0.0359) | | | | |
| UK | KELM | -2.0143 (0.0220) | -1.9951 (0.0230) | -1.7654 (0.0387) | | | |
|  | B-MLP | -2.5639 (0.0052) | -2.3674 (0.0090) | -1.8123 (0.0350) | -1.7741 (0.0380) | | |
|  | MLP | -3.3691 (0.0004) | -3.1231 (0.0009) | -2.1526 (0.0157) | -1.9526 (0.0254) | -1.6979 (0.0448) | |
|  | SARIMAX | -4.2141 (0.0000) | -4.0142 (0.0000) | -3.9896 (0.0000) | -3.6362 (0.0001) | -2.9958 (0.0014) | -2.2036 (0.0138) |
|  | SAKE | -1.7958 (0.0363) | | | | | |
|  | B-KELM | -1.9011 (0.0286) | -1.7896 (0.0368) | | | | |
| Germany | KELM | -2.0034 (0.0226) | -1.9568 (0.0252) | -1.7963 (0.0362) | | | |
|  | B-MLP | -2.4968 (0.0063) | -2.2691 (0.0116) | -1.8856 (0.0297) | -1.7985 (0.0360) | | |
|  | MLP | -3.2103 (0.0007) | -3.1029 (0.0010) | -2.0367 (0.0208) | -2.0014 (0.0227) | -1.7042 (0.0442) | |
|  | SARIMAX | -4.1953 (0.0000) | -4.0183 (0.0000) | -3.8561 (0.0001) | -3.6345 (0.0001) | -3.0152 (0.0013) | -2.3671 (0.0090) |
|  | SAKE | -1.7013 (0.0444) | | | | | |
|  | B-KELM | -1.8354 (0.0332) | -1.7142 (0.0432) | | | | |
| France | KELM | -1.8964 (0.0290) | -1.9648 (0.0247) | -1.7969 (0.0362) | | | |
|  | B-MLP | -2.1243 (0.0168) | -2.0356 (0.0209) | -1.9153 (0.0277) | -1.8015 (0.0358) | | |
|  | MLP | -3.1425 (0.0008) | -3.0142 (0.0013) | -2.0148 (0.0220) | -1.9686 (0.0245) | -1.7126 (0.0434) | |
|  | SARIMAX | -4.0968 (0.0000) | -3.9163 (0.0000) | -3.7853 (0.0001) | -3.5964 (0.0002) | -2.9354 (0.0017) | -2.2012 (0.0139) |

Table 9 The PT test results for different models

| Countries | Horizons | SARIMAX | MLP | B-MLP | KELM | B-KELM | SAKE | B-SAKE |
|---|---|---|---|---|---|---|---|---|
| US | 1-month-ahead | 1.9856 (0.0471) | 2.2013 (0.0277) | 2.9856 (0.0028) | 3.1025 (0.0019) | 3.8842 (0.0001) | 4.3568 (0.0000) | 4.9158 (0.0000) |
| | 3-month-ahead | 1.8335 (0.0667) | 2.0109 (0.0443) | 2.7992 (0.0051) | 2.8519 (0.0043) | 3.6539 (0.0003) | 4.0985 (0.0000) | 4.5985 (0.0000) |
| | 6-month-ahead | 1.2096 (0.2264) | 1.9803 (0.0477) | 2.3981 (0.0165) | 2.4913 (0.0127) | 3.3981 (0.0007) | 3.7251 (0.0002) | 4.0856 (0.0000) |
| UK | 1-month-ahead | 1.9921 (0.0464) | 2.2103 (0.0271) | 2.9936 (0.0028) | 3.1127 (0.0019) | 3.8911 (0.0001) | 4.3602 (0.0000) | 4.9011 (0.0000) |
| | 3-month-ahead | 1.8435 (0.0653) | 2.0127 (0.0441) | 2.8024 (0.0051) | 2.8893 (0.0039) | 3.6694 (0.0002) | 4.1025 (0.0000) | 4.5894 (0.0000) |
| | 6-month-ahead | 1.2109 (0.2259) | 1.9934 (0.0462) | 2.3845 (0.0171) | 2.5005 (0.0124) | 3.4038 (0.0007) | 3.7358 (0.0002) | 4.0952 (0.0000) |
| Germany | 1-month-ahead | 1.9735 (0.0484) | 2.1980 (0.0279) | 2.8913 (0.0038) | 3.0106 (0.0026) | 3.7569 (0.0002) | 4.2011 (0.0000) | 4.7351 (0.0000) |
| | 3-month-ahead | 1.8037 (0.0713) | 2.0014 (0.0453) | 2.7971 (0.0052) | 2.7985 (0.0051) | 3.5050 (0.0005) | 4.0023 (0.0001) | 4.3958 (0.0000) |
| | 6-month-ahead | 1.2003 (0.2300) | 1.9802 (0.0477) | 2.2021 (0.0277) | 2.4801 (0.0131) | 3.3912 (0.0007) | 3.6918 (0.0002) | 3.9965 (0.0001) |
| France | 1-month-ahead | 1.9825 (0.0474) | 2.1833 (0.0290) | 2.7395 (0.0062) | 3.0012 (0.0027) | 3.6028 (0.0003) | 4.2125 (0.0000) | 4.6981 (0.0000) |
| | 3-month-ahead | 1.7953 (0.0726) | 1.9952 (0.0460) | 2.7102 (0.0067) | 2.6528 (0.0080) | 3.4167 (0.0006) | 3.8996 (0.0001) | 4.2896 (0.0000) |
| | 6-month-ahead | 1.1952 (0.2320) | 1.9733 (0.0485) | 2.1537 (0.0313) | 2.3341 (0.0196) | 3.2533 (0.0011) | 3.5853 (0.0003) | 3.9561 (0.0001) |

### 4.5 Summary

In this section, we train the models and conduct numerical experiments employing the multidimensional data related to travel demand, including historical data on tourist arrivals in Beijing from four countries, economic variable data and search intensity index data. The forecasting results of our proposed B-SAKE model and benchmark models are compared through different forecasting schemes. In summary, some interesting implications are obtained as follows.

(1) The proposed B-SAKE model achieves the highest forecast accuracy via MAPE, NRMSE, and DS and outperforms the other baseline models in the DM test and the PT test, followed by other AI models, whereas the SARIMAX model ranks the last.

(2) Although the SARIMAX model can effectively capture the periodicity caused by seasonal factors, it is not applicable to forecasting nonlinear, uncertain and irregular tourist data, while nonlinear AI models have significant advantages.

(3) As an ensemble approach, bagging can effectively increase the data volume available for model training and improve the forecast accuracy through the idea of the model average. Stacked autoencoders are utilized to a construct deep learning network effectively solving the problem of feature learning.

(4) This study analyzes the forecasting performance of the models based on tourist arrival data in Beijing from four countries and comes to the almost consistent conclusions, which illustrates the effectiveness and robustness of this model framework.

### 5. Conclusions and implications

Considering the tourism big data, a bagging-based multivariate ensemble deep learning model, integrating stacked autoencoders and KELM, is proposed for tourism demand forecasting. The data applied in this article include historical data on tourist arrivals in Beijing, economic variable data and search intensity index data. The empirical study results indicate that our proposed B-SAKE model substantially outperforms the benchmark models in different forecasting scheme (onestep-ahead versus multistep-ahead, and in-sample versus out-of-sample). In particular, we analyze the cases of forecasting tourist arrivals from four countries and reach consistent conclusions. Moreover, Bagging and SAE in the ensemble deep learning model framework designed by us effectively solve the overfitting problem and improve the forecasting accuracy by increasing the data volume and improving the feature extraction efficiency respectively.

The ensemble deep learning model we propose can benefit relevant government officials and tourism practitioners. On the one hand, the model has a clear framework and is easy to promote; on the other hand, its powerful prediction performance can help government departments and tourism enterprises to optimize the layout of tourism infrastructure and the allocation of tourism resources, so as to save costs and improve the experience of tourists. In addition, the ensemble deep learning model can also be performed to forecast other complex problems, for instance, transport passenger flow forecasting, air transport demand forecasting, and Precious price forecasting.

The limitation of this study is that other newly developed deep learning models in the tourism forecasting literature don't serve as benchmark models. This is because the details of model construction have not been clearly stated in previous literature, and there are few Shared datasets that contributing to comparability. Fortunately, Y. Zhang et al. (2020) has noticed this issue and release their dataset on GitHub. Additionally, it is worth noting that weather, safety factors, and online comment data have been taken into consideration in the tourism demand forecasting literature (Sohrabi, Raeesi Vanani, Nasiri, & Ghassemi Rudd, 2020). These data can theoretically be incorporated into our proposed B-SAKE model to further improve the forecast accuracy. It is also meaningful to explore the preprocessing of these types of data and the construction of indicators related to tourism demand.

**Conflicts of interest**

The authors declare that they have no conflicts of interest regarding the publication of this study.

**Authors' contributions list**

**Shaolong Sun:** Conceptualization, Methodology, Software, Data Curation, Writing - Review & Editing, Funding acquisition. **Yanzhao Li:** Writing - Original Draft, Visualization. **Ju-e Guo:** Supervision, Funding acquisition. **Shouyang Wang:** Conceptualization, Validation, Funding acquisition.

**Acknowledgements**

This research work was partly supported by the National Natural Science Foundation of China under Grants No. 71988101 and No. 71642006, the Fundamental Research Funds for the Central Universities under Grant No. xpt012020022.